\documentclass[letter]{jpsj2} %% for letters
\title{Evolution of Hall coefficient in two-dimensional heavy fermion CeCoIn$_5$}

\author{Y. \textsc{Nakajima}$^{1}$\thanks{yasuyuki@issp.u-tokyo.ac.jp}, K. \textsc{Izawa}$^{1,2}$, Y. \textsc{Matsuda}$^{1,3}$, K. \textsc{Behnia}$^{1,4}$, H. \textsc{Kontani}$^{5}$, M. \textsc{Hedo}$^{1}$, Y. \textsc{Uwatoko}$^{1}$, T. \textsc{Matsumoto}$^{6}$, H. \textsc{Shishido}$^{7}$, R. \textsc{Settai}$^{7}$, and Y. \textsc{Onuki}$^{7}$}

\inst{$^{1}$Institute for Solid State Physics, University of Tokyo, Kashiwanoha, Kashiwa, Chiba 277-8581, Japan\\
$^{2}$CEA-Grenoble, 38054 Grenoble cedex 9 France\\
$^{3}$Department of Physiscs, Kyoto University, Kyoto 606-8502, Japan\\
$^{4}$Laboratoire de phyque quantique (CNRS), ESPCI, 10 rue de Vauquelin, 75231 Paris, France\\
$^{5}$Department of Physics, Nagoya University, Furo-cho, Chikusa-ku, Nagoya 464-8602, Japan\\
$^{6}$National Institute of Material Science, Sakura, Tsukuba, Ibaraki 305-0003, Japan\\
$^{7}$Graduate School of Science, Osaka University, Toyonaka, Osaka 560-0043, Japan}

\abst{We report on the pressure dependence of the Hall coefficient $R_H$ in quasi-2D heavy fermion CeCoIn$_5$.   At ambient pressure,  below a temperature  associated with the emergence of non-Fermi liquid properties, $R_H$  is anomalously enhanced. We found that the restoration of the Fermi liquid  state with applied pressure leads to a gradual suppression of this dramatic enhancement.  Moreover, the enhancement in  $R_H$ was found to be confined to an intermediate temperature window, where inelastic electron-electron scattering is dominant. Our results strongly support the presence of cold and hot spots on the Fermi surface probably due to anisotropic  scattering by antiferromagnetic fluctuations,  which may also prove relevant for  the debate on the anomalous normal-state properties of high-$T_c$ cuprates.}

\kword{Hall coefficient, pressure, heavy fermion, CeCoIn$_5$, high-$T_c$ cuprates}

\begin{document}
\maketitle

Various transport properties of strongly correlated systems,   including high-$T_c$ cuprates, organics and heavy fermion intermetallics,   exhibit a striking deviation from conventional Fermi liquid behavior.   Clarifying these peculiar transport properties is very important, since they might be the key to elucidating the mechanism of unconventional superconductivity.   Among the various transport properties,  the dc resistivity $\rho_{xx}$ and the Hall coefficient $R_H$,  which are both fundamental parameters, have drawn particular attention.  In fact,  the peculiar temperature and doping dependence of these quantities in  high-$T_c$ cuprates are  striking manifestations of their unconventional normal state properties.  In conventional metals, $\rho_{xx}$ exhibits a $T^2$-dependence and $R_H$ is independent of temperature and reflects the Fermi surface (FS) topology.   In contrast,  in high-$T_c$ cuprates, $\rho_{xx}$ shows a $T$-linear dependence and  $R_H$ exhibits a strong temperature dependence.    Various models to explain the transport scattering time have been proposed, but none of them seem to have captured the entire picture.  Thus the temperature dependence of the resistivity and the Hall effect  remain as one of the most important unresolved questions in  strongly correlated systems  \cite{anderson,chien,pines,millis,kontani,kanki,hussey}. 

The recent discovery of the quasi-2D heavy fermion superconductor CeCoIn$_5$  ($T_c$ = 2.3~K) has aroused great interest in  strongly correlated system \cite{pet}.  CeCoIn$_5$ shares some unconventional properties with high-$T_c$ cuprates.   In the normal state CeCoIn$_5$ exhibits pronounced non-Fermi-liquid behavior related to the antiferromagnetic (AF) quantum critical point (QCP) \cite{sid,NMR,hc}, for instance,   $T$-linear dependent $\rho_{xx}$, strongly $T$-dependent $R_H$ which is nearly inversely proportional to $T$,  and  violation of  Kohler's rule in the magnetoresistance.   All of these anomalous  transport properties have also been reported in high-$T_c$ cuprates \cite{harris}.  Moreover, several measurements indicate that the superconducting symmetry of CeCoIn$_5$ is most likely to be  $d$-wave with nodes perpendicular to the planes, which could be produced by AF spin fluctuations \cite{izawa,mov}.   The non-Fermi to Fermi liquid crossover behavior as a function of pressure has been reported, and  is  reminiscent of a phase diagram with doping in high-$T_c$ cuprates except for the presence of a pseudo gap (see the inset of Fig.~1) \cite{sid,yashima}.  

The making of transport measurements on CeCoIn$_5$ has some advantages compared to other systems.  In cuprates,  chemical doping introduces scattering centers, which strongly influence the transport properties.  In CeCoIn$_5$, on the other hand,  one can tune the electronic structure in a wide range from the non-Fermi to the Fermi liquid state by applying pressure without introducing additional scattering centers.   Moreover, in most  heavy fermion compounds, a large contribution of the skew scattering due to the spin-orbit coupling of $f$-electrons masks the orbital Hall effect.  In contrast,  skew scattering is absent or extremely small  in CeCoIn$_5$ \cite{nakajima,la}.  Thus CeCoIn$_5$ provides a unique opportunity to study non-Fermi liquid behavior in correlated matter and holds promise of bridging our understanding of the heavy fermion materials and cuprates.  

\begin{figure}[t]
\begin{center}
\includegraphics[width=8cm]{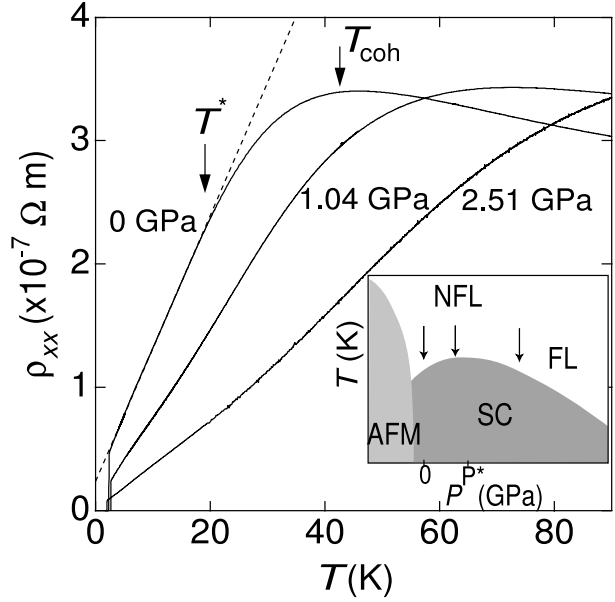}
\end{center}
\caption{(a)Temperature dependence of $\rho_{xx}$ in CeCoIn$_5$ at $P$=0, 1.04 and 2.51~GPa.  At ambient pressure,  $\rho_{xx}$ shows a maximum at $T_{coh} \simeq$45~K and  exhibits a nearly perfect $T$-linear dependence below $T^{\ast}$. Dashed line represents $T$-linear relation of $\rho_{xx}$.   Inset: Schematic $P-T$ phase diagram for CeCoIn$_5$ \cite{sid}.  At $P \sim P^{\ast} =$1.2~GPa, non-Fermi liquid (NFL) to Fermi liquid (FL) crossover occurs and $T_c$ shows a maximum $T_c\simeq$ 2.6~K.  At negative pressure, AF metallic (AFM) state is speculated.  Arrows indicates the pressures applied in this study. }
\end{figure}

In this Letter, we address the issue of the evolution of the Hall coefficient in CeCoIn$_5$ by using pressure as a tuning parameter of the AF quantum criticality.   We report that the prominent $T$-dependence of $R_H$ at ambient pressure is strongly suppressed as the Fermi liquid regime is approached under  pressure.  In addition $R_H$ approaches the Fermi liquid value at very low temperatures.  These results are well explained by the anisotropic scattering rate around the FS.   Several common features in the transport anomaly  in high-$T_c$ cuprates and CeCoIn$_5$ highlight the Hall effect in strongly correlated systems.  

High-quality single crystals of CeCoIn$_5$ and LaCoIn$_5$ were grown by the self-flux method.   The in-plane Hall coefficients were measured for {\boldmath $H$} $\parallel c$.  Hydrostatic pressure up to 2.51 GPa was generated in a piston-cylinder type high pressure cell with oil as a transmitting fluid (Daphene 7373).  The pressure inside the cell was determined by the superconducting transition temperature of Pb.  

Figure 1 depicts the temperature dependence of  $\rho_{xx}$ for CeCoIn$_5$.    The transition temperatures are 2.3, 2.6, and 1.9~K, at $P$=0, 1.04, and 2.51~GPa, respectively, which are close to the reported values \cite{sid}.    Associated with an incoherent-coherent crossover,  $\rho_{xx}$ shows  maxima at $T_{coh}\sim$ 45, 70, and 150  at $P$=0, 1.04, and 2.51~GPa, respectively.      Below $T^{\ast}\simeq$ 20~K at ambient pressure $\rho_{xx}$ displays an almost perfect $T$-linear dependence down to $T_c$.   An increase in pressure leads to an increase in $T_{coh}$ and  a decrease in $\rho_{xx}$  due to enhancement of the hybridization between conduction electrons and $f$-electrons.   At $P$ = 2.51~GPa,  an apparent deviation from $T$-linear dependence is observed in $\rho_{xx}$; the overall $T$-dependence above $T_c$ to $T\sim$ 60~K is well fitted as $\rho_{xx}\propto T^{\alpha}$ with $\alpha \approx$ 1.2, indicating that the system approaches  the Fermi liquid regime.  The inset of Fig.~1 illustrates the schematic temperaure-pressure phase diagram for CeCoIn$_5$ reported in Ref.\cite{sid}.  Because of its relationship to CeRhIn$_5$,  the AF metallic phase of CeCoIn$_5$ is speculated to occur at a slightly negative pressure.   At $P^{\ast} \sim$ 1.2~GPa the non-Fermi liquid to Fermi liquid crossover occurs and $T_c$ shows a maximum ($T_c\simeq$ 2.6~K).   %As reported in Ref., $\rho_{xx}$ at $P$=2.51~GPa exhibits $T^2$-dependence  in a narrow  temperature range just above $T_c$ to $T_{FL}\simeq$2.7~K, as shown in the inset of Fig.~1 (c).  

\begin{figure}[t]
\begin{center}
\includegraphics[width=8cm]{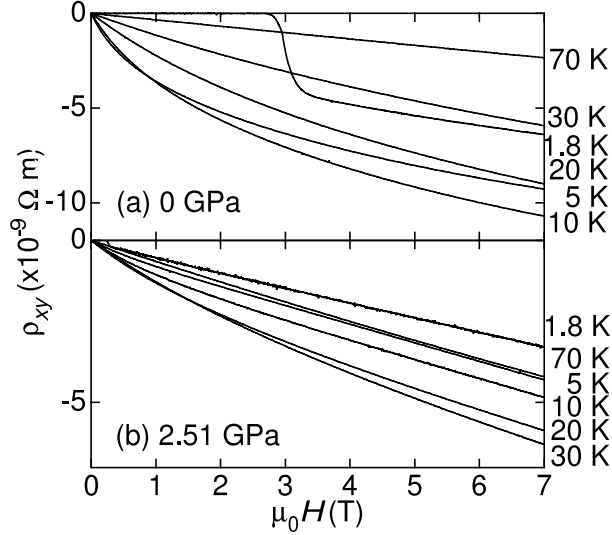}
\end{center}
\caption{Hall resistivity $\rho_{xy}$ of CeCoIn$_5$ as a function of $H$ at (a) 0 and (b) 2.51~GPa for temperatures from 1.8~K to 70~K.}
\end{figure}

Figures 2(a) and (b) show the $H$-dependence of the Hall resistivity $\rho_{xy}$ at $P$ = 0 and 2.51~GPa, respectively.  The Hall sign is negative and $\rho_{xy}$ exhibits a non-linear $H$-dependence, which is more pronounced at ambient pressure.   We here define $R_H$ as the field derivative of $\rho_{xy}$, $R_H\equiv\frac{{\rm d}\rho_{xy}}{{\rm d}H}$.  Figure 3 depicts the Hall coefficient at the zero field limit, together with the same data for LaCoIn$_5$ with no $f$-electron, $R_H^{La}$.  We note that  $\rho_{xy}$ for LaCoIn$_5$ shows nearly perfect $H$-linear dependence.  With decreasing $T$, $R_H^{La}$ shows a slight increase and the amplitude at low temperature is $-5.5 \times 10^{-10}$ m$^3$/C, which is nearly half  the Hall coefficient expected for one electron per unit cell.

\begin{figure}[t]
\begin{center}
\includegraphics[width=8cm]{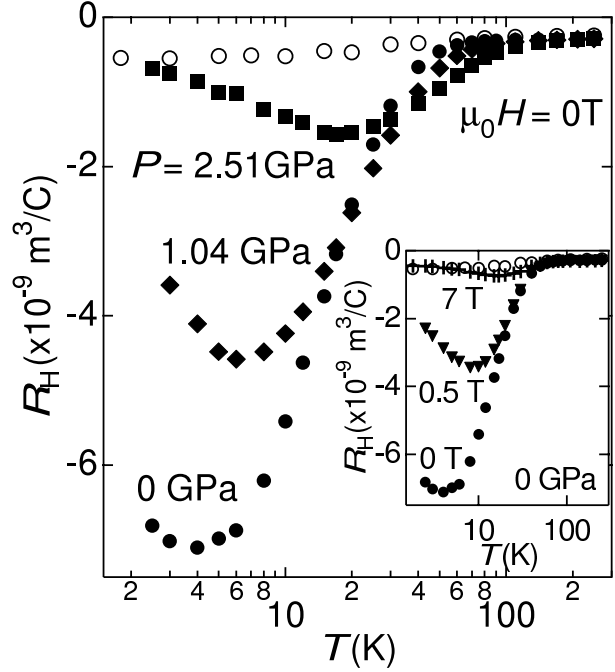}
\end{center}
\caption{Temperature dependence of $R_H$ for CeCoIn$_5$ at $P$ = 0 ($\bullet$), 1.04 ($\blacklozenge$),  and 2.51~GPa ($\blacksquare$) at  the zero field limit, $\lim_{H\rightarrow0}\frac{{\rm d}\rho_{xy}}{{\rm d}H}$, and for LaCoIn$_5$,  $R_H^{La}$ ($\circ$).  Inset:  The same data for CeCoIn$_5$ at  ambient pressure at  0~T ($\bullet$), 0.5~T ($\blacktriangledown$),  and  7~T ($+$),  defined as ${\rm d}\rho_{xy}/{\rm d}H$,  and  for LaCoIn$_5$ ($\circ$). }
\end{figure}

The temperature dependence of the Hall effect is closely correlated with that of the resistivity.  At high temperatures $T>T_{coh}$,  $R_{H}$ of CeCoIn$_5$ at all pressures is nearly  independent of temperature and coincides well with $R_H^{La}$.  This is consistent with  band structure calculations which predict that the electronic structure of CeCoIn$_5$ is similar to that of LaCoIn$_5$ \cite{shishido,harima}.   We therefore assume that  $R_H^{La}$ is close to the Hall coefficient of CeCoIn$_5$ in the Fermi liquid limit, though $R_H^{La}$ should not perfectly coincide with $R_H$ of CeCoIn$_5$ because below $T_{coh}$ the carrier number of CeCoIn$_5$ is larger than that of LaCoIn$_5$  due to the itinerant $f$-electron.   Above $T_{coh}$, phonon and incoherent Kondo scattering determine the transport properties.  Below $T_{coh}$,   $R_H$ exhibits a striking departure from $R_H^{La}$.  For all pressures, as the temperature is lowered,  $R_H$ decreases rapidly.  Further decreases of temperature increase $R_H$ after showing minima at around 4, 6, and 20~K for $P$ = 0, 1.04, and 2.51~GPa, respectively.   At ambient pressure, the absolute value of $R_H$ at the minimum is enhanced to nearly 30 times  that at high temperature above $T_{coh}$.  This enhancement is dramatically suppressed by  pressure.  Interestingly,  $R_H$ at $P$ = 2.51~GPa approaches  $R_H^{La}$ at very low temperatures. We will discuss this behavior later.   Thus there are three distinct temperature regions in the Hall effect,  denoted as (I),  (II), and (III)  as illustrated in Fig.~4(a).  

\begin{figure}[t]
\begin{center}
\includegraphics[width=8cm]{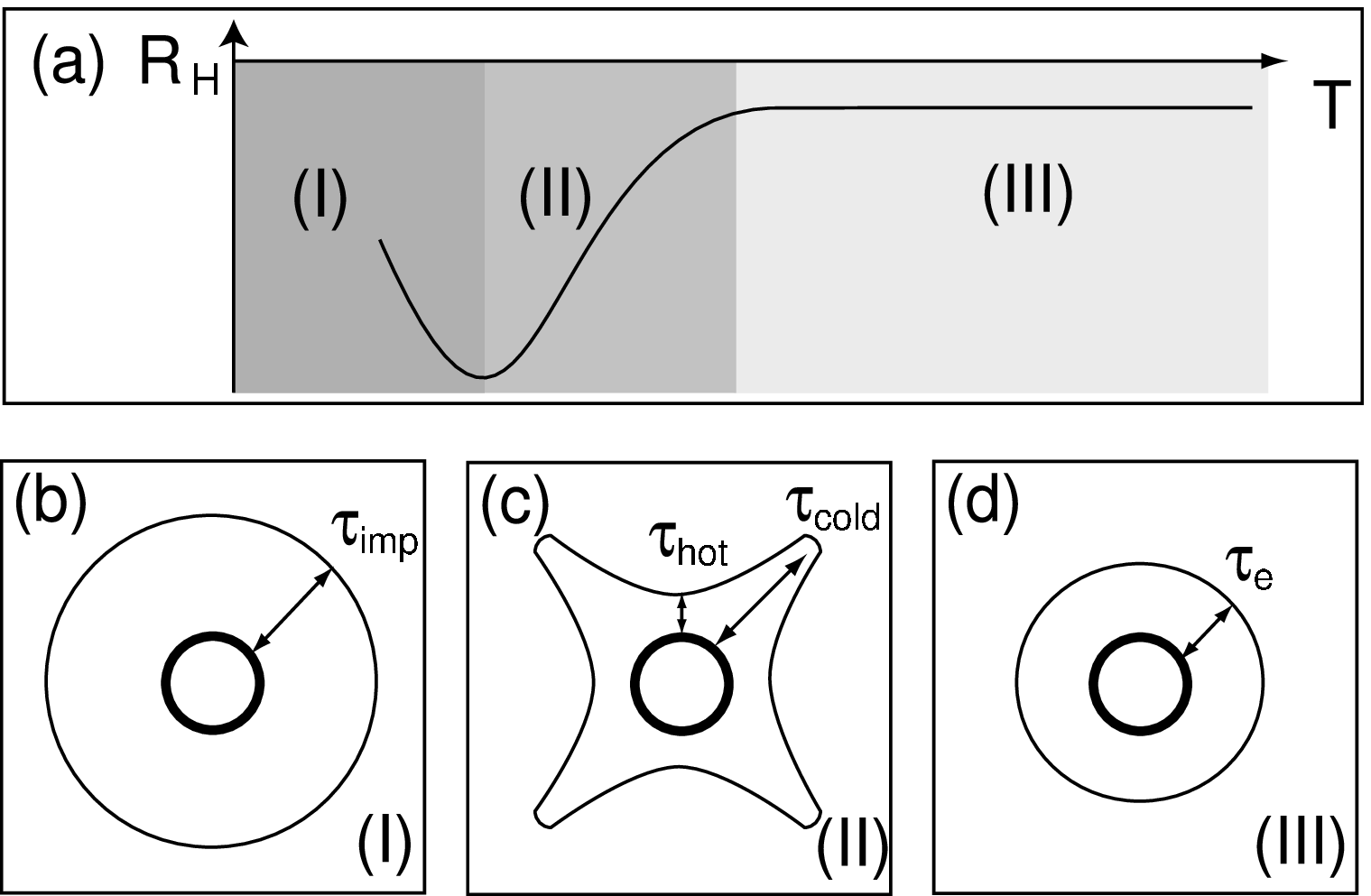}
\end{center}
\caption{(a) Schematic figure of the $T$-dependence of $R_H$ for CeCoIn$_5$.  There are three distinct $T$-regions: High $T$-region (III), where $R_H$ is nearly $T$-independent; intermediate $T$-region (II), where $R_H$ exhibits a rapid decrease; and low $T$-region (I), where $R_H$ increases.  (b), (c), and (d) show the anisotropy of the  scattering time $\tau$($\phi$)  (thin solid lines) around the Fermi surface (solid lines) at region-(I), (II), and (III), respectively.  In region-(II), the scattering time is highly anisotropic,  possibly due to the AF spin fluctuation.  On the other hand,  the scattering time is isotropic in region-(I) where impurity scattering dominates and in region-(III) where  phonon and incoherent Kondo scattering determine the elastic scattering time $\tau_e$.   
}
\end{figure}

It has been argued that in high-$T_c$ cuprates the Hall problem can be simplified when analyzed in terms of $\cot \Theta_H$,  where $\Theta_H\equiv  \tan^{-1} \rho_{xy}/\rho_{xx}$ is the Hall angle \cite{anderson,chien,hussey}.  A very recent systematic study of $R_H$ on La$_{2-x}$Sr$_x$CuO$_4$ single crystals over a wide doping range has shown that $T^2$-law of $\cot \Theta_H$ holds well from the underdoped region up to optimal doping, but gradually breaks down,   showing a negative curvature when the system is overdoped \cite{ando}.   We here examine the pressure dependence of  $\cot \Theta_H$ for CeCoIn$_5$.  Figure 5 depicts $\cot \Theta_H$ as a function of  $(T/T_{coh})^2$.   The slope of $\cot\Theta_H$ increases with pressure.   Moreover, while $\cot \Theta_H$ at $P$ = 0 and 1.04~GPa show a $T^2$-dependence, $\cot \Theta_H$  deviates from the $T^2$-law with a negative curvature at $P$ = 2.51~GPa.  The Hall angle in CeCoIn$_5$ again bears a resemblance to that in high-$T_c$ cuprates.  

The striking similarity  in $\rho_{xx}$, $R_H$, and $\cot\Theta_H$ between high-$T_c$ cuprates and CeCoIn$_5$ lead us to consider that the anomalous transport properties in both systems have a similar origin.   It should be noted that in high-$T_c$ cuprates the presence of the pseudogap in the underdoped regime reduces both $R_H$ and $\rho_{xx}$, while the presence of  the pseudogap in CeCoIn$_5$ is controversial.  

%We here comment on the recent argument of $R_H$ in CeCoIn$_5$ discussed in terms of two-fluid Kondo model , in which  the Kondo lattice system is divided into coherent and incoherent part.   It has been proposed that $R_H$ is expressed as $R_H=\alpha_f R_H^{La}$, where $\alpha_f$ is $f$-electron weighting function reflecting coherent part  \cite{lahall}.  However, if we apply this model to $R_H$ at 2.51GPa,  $\alpha_f$ is close to unity at low temperatures, indicating no coherent part.   This contradicts the itinerant $f$-electrons in this temperature regime.   Therefore the two-fluid Kondo scenario appears to be inconsistent with the present results.

There are two broad  approaches to understanding the transport properties in high-$T_c$ cuprates:  the non-Fermi liquid, in which the transport carriers are exotic objects, spinons and holons \cite{anderson,chien}, and the Fermi liquid, in which the carriers are electrons \cite{pines,millis,kontani,kanki,hussey}.     The former model involves two distinct relaxation times for momentum changes parallel and perpendicular to the FS.    It has been argued that the observation of $T^2$-dependence of  $\cot \Theta_H$ along with the $T$-linear $\rho_{xx}$ implies the existence of different longitudinal and Hall relaxation rates that changes as $\sim$$T$ and $\sim$$T^2$, respectively.  However this idea of scattering rate separation has not yet gained a complete consensus.   Moreover it seems unlikely that the spin-charge separation discussed in relation to cuprates occurs in CeCoIn$_5$.    

In the Fermi liquid approach, an anomalous scattering mechanism is invoked \cite{pines,millis,kontani,kanki,hussey}.  The central concept behind this is the ''hot spot", a small region on the FS where the electron lifetime is unusually short, e.g., due to  electron-AF spin fluctuation scattering (see Fig.~4(c)).   According to Ref. \cite{pines},  $R_H$ can be written as $R_H=\frac{1}{en}\frac{1+r}{2\sqrt{r}}$,  where $n$ is the career density and $r$ is the anisotropy ratio of the scattering time at hot and cold (non-hot) spots, $r \equiv \tau_{cold}/\tau_{hot}$.  However, applying this theory to CeCoIn$_5$ gives an extremely large $r$ at  ambient pressure $r\sim$ 2500.   Such a large $r$ would produce a large magnetoresistance \cite{millis}, $\Delta \rho_{xx}/\rho_{xx}\sim 10r\tan^2 \Theta_H > 100$ at 1~T, but such a phenomenon has never been observed \cite{nakajima}.  

Recently it has been pointed out that the above Fermi liquid approach based on the Boltzmann approximation should be modified due to the backflow effect arising from strong AF fluctuations \cite{kontani,kanki}.   Backflow is a polarized current caused by a quasiparticle excitation inherent in  Fermi liquids.  Backflow caused the Hall coefficient to strongly deviate from the Boltzmann value derived from the curvature of the FS, because the total current ${\bf J}_{\bf k}$ is not parallel to the Fermi velocity ${\bf v}_{\bf k}$ and is not perpendicular to the FS.   As a result, $R_H$ behaves as
\begin{equation}
	R_H \sim \pm \xi_{AF}^2,
\end{equation}
where $\xi_{AF}$ is the AF correlation length.   Near the AF QCP, $\xi_{AF}$ depends on $T$ as $\xi_{ AF}^2 \propto 1/(T+\theta)$, where $\theta$ is the Weiss temperature.  Various transport properties under a magnetic field are dominated by $\xi_{AF}$ \cite{kontani2}.  On the other hand, since  $\rho_{xx}$ is governed by the cold spot (Fig.~4(c)),
\begin{equation}
	\rho_{xx}\sim \tau_{cold}^{-1}
\end{equation}
and $\tau_{cold}^{-1}\propto T$ in  AF fluctuation theory \cite{pines,kontani}.   Hence we obtain $\rho_{xx} \propto T$, $R_{H}\propto T^{-1}$, and $\cot \Theta_H \propto T^2$, which well reproduce the data at $P$ = 0 and 1.05~GPa.  Since  pressure reduces $\xi_{AF}$ by reducing the AF fluctuation,   the amplitude of $R_H$ is reduced with increased pressure,  consistent with experiments.     Further support for this idea is shown in the inset of Fig.~3 depicting $R_H={\rm d}\rho_{xy} /{\rm d} H$ in a magnetic field at ambient pressure.   The amplitude of $R_H$ is strongly suppressed by a magnetic field \cite{nakajima,la}.  At $H$ = 7~T, $R_H$ is very close to $R_H^{La}$.   Since the AF fluctuations are also suppressed by a magnetic field,  the suppression of $R_H$ confirms the idea that AF  fluctuations play an important role in the Hall effect.  

\begin{figure}[t]
\begin{center}
\includegraphics[width=8cm]{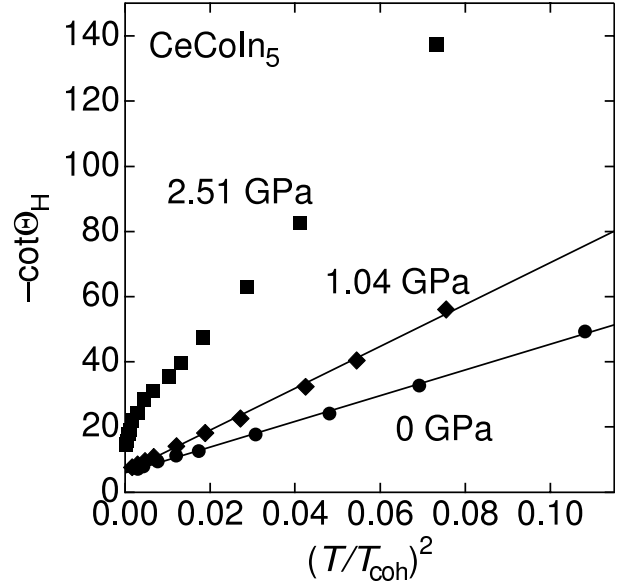}
\end{center}
\caption{$-\cot \Theta_H$ versus $T/T_{coh}$ for CeCoIn$_5$ at $P$ = 0 ($\bullet$), 1.04 ($\blacklozenge$) and 2.51~GPa ($\blacksquare$).  The solid thin lines represent the $T^2$-dependence of $\cot \Theta_H$.}
\end{figure}

We finally discuss  the Hall effect at low temperatures (region-(I) in Fig.~4).   The most remarkable feature at low temperatures is that  $R_H$ at $P$ = 2.51~GPa approaches  $R_H^{La}$.  This indicates that the Hall coefficient, which exhibits a non-Fermi liquid behavior at higher temperatures,  approaches  the Fermi liquid value.    We stress that these behaviors of the Hall coefficient again support  the  AF  fluctuation theory.  According to Kubo formula, the scattering time, $\tau$, can be described by an isotropic impurity part and  anisotropic inelastic part,
\begin{equation}
\tau(\phi)=\frac{1}{\tau_{imp}^{-1}+\tau_{inel}(\phi)^{-1}}. 
\end{equation}
In  region-(II) where inelastic scattering due to AF fluctuation dominates,  $\rho_{xx}$ and $R_H$ are determined by Eqs. (1) and (2).  However, as the temperature is lowered, isotropic impurity scattering becomes important.  As a result, the influence of  AF fluctuation on $R_H$ is  reduced and  hence the scattering becomes isotropic and the backflow vanishes, as shown in Fig.~4(b).   Thus $R_H$ is expected to approach the Fermi liquid value.  At $P$ = 0 and 1.04~GPa, $R_H$ does not recover to $R_H^{La}$ in region-(I), though a slight increase of $R_H$ is observed.  This is because the system undergoes a superconducting transition from the normal state with non-Fermi liquid properties.   These results lead us to conclude that  anisotropic inelastic scattering, most likely arising from AF spin fluctuation, plays a crucial role in the  Hall effect.   It may also be a cause of the  large Nernst effect observed below $T_{coh}$ at ambient pressure in CeCoIn$_5$ \cite{bel}.  The giant Nernst effect which, like the anomalously large Hall coefficient, gradually fades with the application of a magnetic field, may reflect a large energy-dependence of the scattering time and  highly anisotropic backflow due to the presence of AF fluctuations \cite{kontani2}.

To conclude, we have measured the Hall effect of CeCoIn$_5$ under pressure, ranging from the Fermi liquid to non-Fermi liquid regime. The remarkable enhancement of the Hall coefficient observed in the non-Fermi liquid regime is dramatically suppressed as the Fermi liquid regime is approached.  At low temperatures the Hall coefficient  approaches the Fermi liquid value.  These results provide strong evidence that hot and cold spots around the Fermi surface, most likely produced by AF fluctuation, are important for understanding the anomalous transport properties.  Some of these unusual  transport properties in CeCoIn$_5$ bear a striking resemblance to the normal-state properties of high-$T_c$ cuprates, indicating  universal transport properties of strongly correlated systems. 

We thank N.~E.~Hussey, T.~Shibauchi, and K.~Yamada for their valuable discussions. This work was partly supported by a Grant-in-Aid for Scientific Reserch from the Ministry of Education, Culture, Sports, Science and Technology. One of the authors (Y.N.) was supported by the Research Fellowships of the Japan Society for the Promotion of Science for Young Scientists.

\end{document}